# PQStream: A Data Stream Architecture for Electrical Power Quality

Dilek Küçük[1], Tolga İnan[1], Burak Boyrazoğlu[1,2], Serkan Buhan[1,3],
Özgül Salor[1], Işık Çadırcı[1,3], Muammer Ermiş[2]

[1] TÜBİTAK – Uzay, Power Quality Group, Ankara – Turkey
{dilek.kucuk, tolga.inan, burak.boyrazoglu, serkan.buhan, ozgul.salor}@uzay.tubitak.gov.tr
[2] METU, Electrical and Electronics Eng. Dept., Ankara – Turkey
ermis@metu.edu.tr
[3] Hacettepe University, Electrical and Electronics Eng. Dept., Ankara – Turkey
cadirci@ee.hacettepe.edu.tr

**Abstract.** In this paper, a data stream architecture is presented for electrical power quality (PQ) which is called PQStream. PQStream is developed to process and manage time-evolving data coming from the country-wide mobile measurements of electrical PQ parameters of the Turkish Electricity Transmission System. It is a full-fledged system with a data measurement module which carries out processing of continuous PQ data, a stream database which stores the output of the measurement module, and finally a Graphical User Interface for retrospective analysis of the PQ data stored in the stream database. The presented model is deployed and is available to PQ experts, academicians and researchers of the area. As further studies, data mining methods such as classification and clustering algorithms will be applied in order to deduce useful PQ information from this database of PQ data.

## 1 Introduction

The proliferation of time-involving and data-intensive applications such as sensor networks, network traffic monitoring systems and financial applications led to the emergence of data stream models and issues related to the management of these models as well. Considerable research have been carried out on data streams including the studies on data stream management systems such as STREAM [1], those ones on tracking cross-correlation in data streams [2, 3], studies on mining data streams such as StreamCube [4], and finally those studies that present real world applications of data streams such as GigaScope [5]. We refer interested readers to [6, 7] for in-depth surveys of the literature on data streams.

In this paper, a data stream architecture is presented, which is called PQStream, for processing and managing electrical power quality (PQ) data. The feasibility and effectiveness of the proposed architecture is demonstrated on the PQ data obtained by a mobile PQ measurement system [8] monitoring the transformer substations of the

Turkish Electricity Transmission System where this measurement system is also part of the PQStream architecture.

The rest of the paper is organized as follows: Section 2 presents a brief review of electrical power quality and power quality parameters. In Section 3, general PQStream architecture and its main components are described in detail with its application on field PQ data. Finally, conclusions and future research directions are presented in Section 4.

## 2   Electrical Power Quality Parameters

Electrical power is one of the most essential items used by commerce and industry today. It is an unusual commodity because it is required as a continuous flow - it cannot be conveniently stored in quantity - and it cannot be subject to quality assurance checks before it is used [9]. The reliability of the supply must be known and the resilience of the process to variations must be understood. In reality, of course, electricity is very different from any other product – it is generated far from the point of use, is fed to the grid together with the output of many other generators and arrives at the point of use via several transformers and many kilometers of overhead and possibly underground cabling. Where the industry has been privatized, these network assets will be owned, managed and maintained by a number of different organizations. Hence assuring the quality of delivered power at the point of use is no easy task.

Consumers of electricity are being increasingly affected by PQ problems due to augmentation of disturbing loads in electric power systems. Throughout the study, PQ parameters given at IEC 61000-4-30 [10] are used. In the following subsections, basic electric power definitions and brief descriptions of some PQ parameters can be found. At the end of this section, event types such as sag, swell, unbalance and interrupt are explained as well.

### 2.1  Power

Electric power is defined as the amount of work done by an electric current, or the rate at which electrical energy is transferred. In alternating current circuits, energy storage elements such as inductance and capacitance may result in periodic reversals of the direction of energy flow. The portion of power flow that averaged over a complete cycle of the AC waveform, which results in net transfer of energy in one direction is known as real power. That portion of power flow due to stored energy that returns to the source in each cycle is known as reactive power. The relationship between real power, reactive power and apparent power can be expressed by representing the quantities as vectors (Fig.1). The apparent power vector is the hypotenuse of a right triangle formed by connecting the real and reactive power vectors [11].

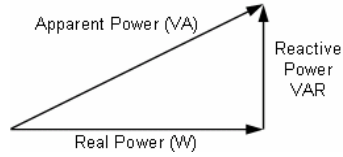

**Fig. 1.** Relation between Real, Reactive, and Apparent Powers

### 2.2 Demand

Electric power demand is directly proportional to the current demand of consumer. Hence consumer's power demand profile may obtained by sampling the current demand values. Power demand is mainly characterized by fundamental harmonic component of the load current.

### 2.3 Voltage and Current RMS

RMS (Root-Mean-Square) value is defined to be square root of the arithmetic mean of the squares of the instantaneous values of a quantity taken over a specified time interval. The average power consumed by a sinusoidally driven linear two-terminal electrical device is a function of the RMS values of the voltage across the terminals and the current passing through the device, and of the phase angle between the voltage and current sinusoids.

### 2.4 Frequency

The mains frequency is the frequency at which alternating current is transmitted from a power plant to the end user. In most parts of the world, it is typically 50 or 60 Hz. Mains frequency is fixed to 50 Hz for the Turkish Electricity Transmission System. However due to practical load demand variations, supply frequency appears to be in a frequency band rather than having a constant value. Hence, mains frequency turns out to be an important PQ parameter indicating the frequency stability of the particular utility grid.

### 2.5 Harmonics

Ideally, voltage and current waveforms are perfect sinusoids. However, as reported in [12], because of the increased popularity of electronic and other non-linear loads, these waveforms often become distorted. This deviation from a perfect sine wave can be represented by harmonics—sinusoidal components having a frequency that is an integral multiple of the fundamental frequency (Fig. 2). Thus, a pure voltage or current sine wave has no distortion and no harmonics, and a non-sinusoidal wave has distortion and harmonics. To quantify the distortion, the term total harmonic distortion (THD) is used. The term expresses the distortion as a percentage of the fundamental (pure sine) of voltage and current waveforms.

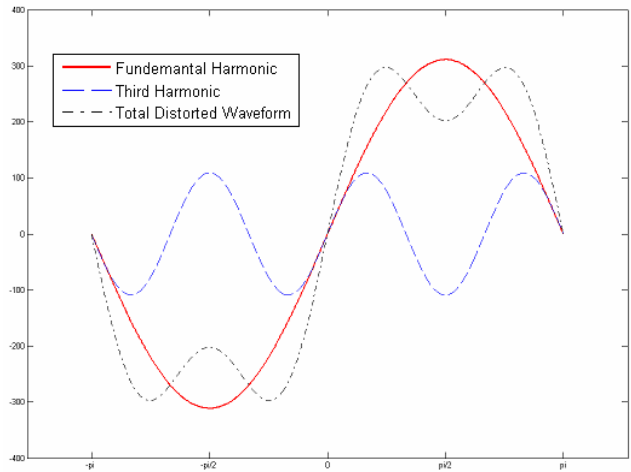

**Fig. 2.** Distorted Waveform Composed of Fundamental and 3rd Harmonic

### 2.6 Flicker

The power supply network voltage varies over time due to perturbations that occur in the processes of electricity generation, transmission and distribution. Interaction of electrical loads with the network causes further deterioration of the electrical PQ. High power loads that draw fluctuating current, such as large motor drives and arc furnaces, cause low frequency cyclic voltage variations that result in flickering of light sources which can cause significant physiological discomfort, physical and psychological tiredness, and even pathological effects for human beings. Hence, flicker is quantified based on models of light sources and human sensation [13].

### 2.7 Events

Voltage sag, swell, unbalance and interruption are detected as PQ events throughout the measurements. These events are briefly described below [10]:

- *Voltage Sag:* Sag indicates an under-voltage situation. On poly-phase systems, voltage sag begins when the voltage of one or more channels is below a sag threshold (%85 of nominal) and ends when voltage on all measured channels is equal to or above the sag threshold plus the hysteresis voltage.

- *Voltage Swell:* Swell indicates an over-voltage situation. On poly-phase systems, a swell begins when the voltage of one or more channel rises above the swell threshold (%110 of nominal) and ends when the voltage on all measured channels is equal to or below the swell threshold minus the hysteresis voltage.

- *Unbalance:* Channels of poly-phase systems should have sinusoidal voltages having same amplitude. Balanced three-phase system should have the same amplitude on all phases. Unbalance is a measure which indicates how much the amplitude of phases different from each other.

- *Interruption:* On poly-phase systems, a voltage interruption begins when the voltage of all channels is below the voltage interruption threshold (%5 of nominal) and ends when the voltage of any one channel is equal to or greater than the voltage interruption threshold plus the hysteresis.

## 3  PQStream Architecture

PQStream is an architecture offered for efficient processing and management of PQ data. PQ data has an inherent time-dependency and this data when measured at relevant frequencies requires almost unbounded storage and processing capabilities compared to other data types stored in conventional relational database management systems.

We have used PQStream architecture for mobile measurements of PQ data in the Turkish Electricity Transmission System and describe the architecture accordingly, yet it is a generic architecture and can be used for managing any PQ data acquired in other means with little or no customization.

An abstract representation of PQStream presented in Fig. 3. In the following subsections, we firstly introduce the mobile PQ measurements application which corresponds to the PQ measurement module in Fig. 3 and describe how the resulting data is transferred to the stream database. Following this, we describe the stream database with its conceptual data model and finally the PQStream GUI.

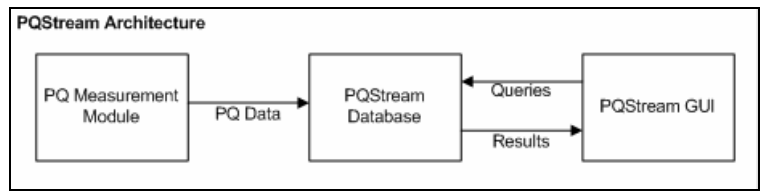

**Fig. 3.** Abstract Representation of PQStream Architecture

### 3.1 PQ Data Measurement Module

Mobile PQ measurements in the Turkish Electricity Transmission System are carried out for a period of seven consecutive days for each measurement point. Measurement points are feeders and busbars in the transformer substations. At the time this paper is written, measurements have been completed for 144 bus-bars, 205 feeders, 59 transformer substations all over the country.

Mobile measurement program is developed in LabView development environment [14] using its proprietary visual programming language called G where the sampling frequency of the program is 3200 Hz, that is, it acquires 3200 raw samples per second for each PQ parameter. The program calculates and outputs the averages corresponding to the PQ parameters according the PQ standards [10] in an online fashion. It outputs raw PQ data as well in case of events as will be clarified in the upcoming paragraphs.

Output traffic load of the PQStream measurement module is presented in Table 1 for each PQ data measurement point. Storage and processing requirements of PQStream database could be estimated using the total bit rate values in this table and number of measurement points.

**Table 1.** Outgoing Data Traffic of PQStream Measurement Module (Based on Mobile PQ Measurements of the Turkish Electricity Transmission System).

| Parameter | Precision | Update Rate (Averaging Interval) | Three Phase | Average PQ Data Bit Rate (bps) |
|---|---|---|---|---|
| Active Power | Double | every second | Yes | 192 |
| Reactive Power | Double | every second | Yes | 192 |
| Apparent Power | Double | every second | Yes | 192 |
| Power Factor | Double | every second | Yes | 192 |
| 33 Voltage Harmonics | Double | every 3 secs. | Yes | 2.112 |
| 33 Current Harmonics | Double | every 3 secs. | Yes | 2.112 |
| RMS Current and Voltage | Double | every 0.2 secs. | Yes | 1.920 |
| Event Length | Integer | variable | No | 4 |
| Event Type | String | variable | No | 10 |
| Event Raw Current[1] Data | Double | variable | Yes | 614.400 |
| Event Raw Voltage Data | Double | variable | Yes | 614.400 |
| Short Term Flicker | Double | every 10 mins. | Yes | 0,32 |
| Demand | Double | every 15 mins. | Yes | 0,213 |
| Frequency | Double | every second | No | 64 |
| | | | Total (with Events) | 1.235.790,533 |
| | | | Total (without Events) | 6.990,533 |

---

[1] 614400 bps = 3200 samples/sec*8 bytes/sample*8 bits/byte*1 sample/phase*3 phase

The output of the program is a set of directories and files for each PQ parameter where exact directory structure of this output is presented in Fig. 4.

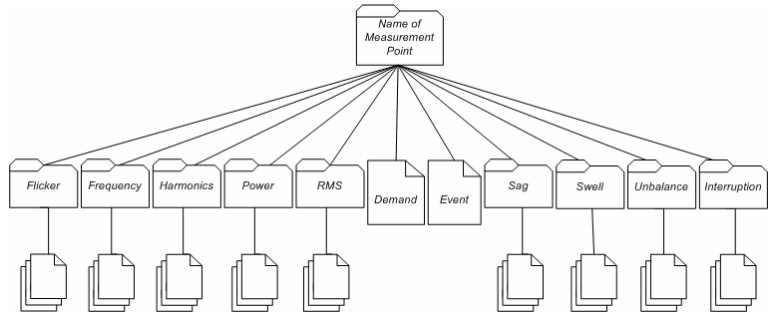

**Fig. 4.** Directory Structure of Mobile PQ Measurements

In order to store and manage this PQ measurement data in a stream database, a daemon program executes on the data in order to transfer them to the database. The program simply reads the measurement files for PQ parameters and inserts each sample in these files to the corresponding tables in the stream database.

However, not all values in all files in Fig. 4 are directly stored in the database due to space and processing limitations. The files under the directories of *Sag, Swell, Unbalance*, and *Interruption* include *raw data* corresponding to the actual samples during the entire period of each event and are simply stored as compressed files in the file system in a specific directory layout instead of storing their contents in the database. Only absolute paths of these files are stored in the database. *Event* measurement file in Fig. 4 has an entry for each of these events so that this information will be available through the database and if the actual *raw data* is required for an event, it will be provided to the user as a file. We refer interested readers to [8] for an in-depth description of the mobile PQ measurements application.

### 3.2 Data Stream Model for PQ Data

The output of the PQStream data measurement module, which corresponds to the computed averages of PQ parameters according to the averaging intervals provided in Table 1, should be effectively stored in a database for retrospective analysis of the PQ data. For this purpose, we have proposed a conceptual data model for PQ data and presented this model as a Unified Modeling Language (UML) class diagram in Fig. 5.

PQStream database is constructed by implementing each of the classes in Fig. 5 as tables of a database using open-source object relational PostgreSQL as the backend database system. These classes are briefly described below:

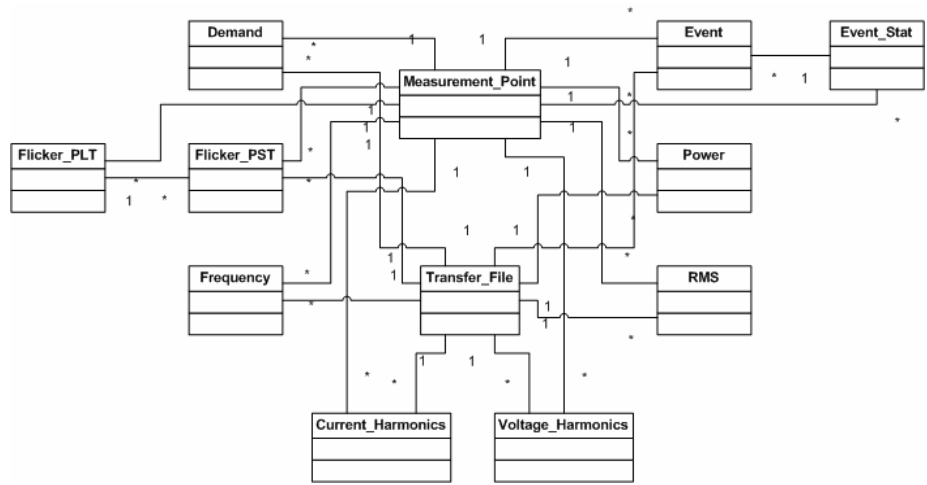

**Fig. 5.** Conceptual Data Model for PQ Data Represented with a UML Class Diagram

- *Measurement_Point* class holds information about the busbars or feeders (measurement points) where PQ measurements take place. The attributes of this class are crucial since they could be used to group stream data (hence will be in the *group by* clause when the required data is represented as a Structured Query Language (SQL) query or a query in one of the other languages based on SQL). Some of the most significant attributes of this class are *load_type* (which can take on one of the values of *Heavy Industry, Industry+Urban,* and *Urban Only*), *city_name, region_name,* and *voltage_level*.

- *Transfer_File* class is for holding information related to the transfer file and actual data transfer time. Since each PQ measurement sample has a corresponding timestamp; this value must also be stored in the stream database. But, frequency of each PQ parameter is determined according to PQ standards (provided as averaging intervals in Table 1), that is, duration between consecutive timestamps are known in advance, hence we use an attribute called *measurement_date* for each *Transfer_File* instance to represent the measurement time of the last sample in that file so that the timestamps for the remaining samples could be determined according to the PQ parameter type.

- *Event* class is used to model an entry for each and every event that occurred during the entire measurement period. The attributes of this class include *event_type* (one of *sag, swell, interruption* or *unbalance*), *event_starting_time*, *event_ending_time,* and *event_size_in_samples*. Although *raw data* corresponding to each event occurrence is also stored in the directory structure in Fig. 4 as a file for each event type under corresponding directories, they are not individually modeled in the conceptual design due to space and processing limitations (the number of event occurrences is bounded only by the total measurement period) and we store these files in compressed form in a certain directory structure for each measurement point as explained at the end of Section 3.1. The absolute path of each of these raw data files is modeled with the *file_path* attribute of the *Event* class.

- *Event_Stat* is a class introduced for efficiency reasons. It is used to model a summary of the events occurred in a measurement point. The attributes of *Event_Stat* include *event_count, sag_count, swell_count, interruption_count* and *unbalance_count*. With this class implemented as a database table, most of the aggregation queries on PQStream will be faster (since they will scan *Event_Stat* table instead of the larger *Event* table).

- Among the remaining classes, *Flicker_PST* models short term, and *Fliker_PLT* models long term flicker measurements. Long term flicker ($P_{lt}$) is calculated from short term flicker ($P_{st}$) with the formula (1) taking N=12 where $P_{sti}$ (i=1..N) are consecutive values of $P_{st}$. As their names imply, the classes *Demand*, *Frequency*, *RMS* and *Power* are for modeling respective PQ parameters.

$$P_{lt} = \sqrt[3]{\frac{\sum_{i=1}^{N} P_{sti}^3}{N}} \qquad (1)$$

### 3.3 PQStream Graphical User Interface (GUI)

A user-friendly interface is essential for the effective querying of the presented PQStream database. Furthermore, this interface should provide high-quality visualization facilities to its users since graphics is probably the best way to present PQ data.

For this purpose, a GUI for PQStream has been developed using Java programming language, with its Swing Application Programming Interface (API), in Eclipse development environment. The characteristics of PQStream GUI are summarized below:

- It enables its users to query each of the PQ parameters and results can be represented using different visualization options such as tables, bar/pie charts, or time-series graphics. Graphics facilities are implemented using the open-source JFreeChart API [15]. In Fig. 6, flicker panel of PQStream GUI is presented where the query provided through the GUI results in the graphical representation of short term flicker for a measurement point, namely, *EZİNE TM 154/34.5 KV TRAFO PRİMERİ*.

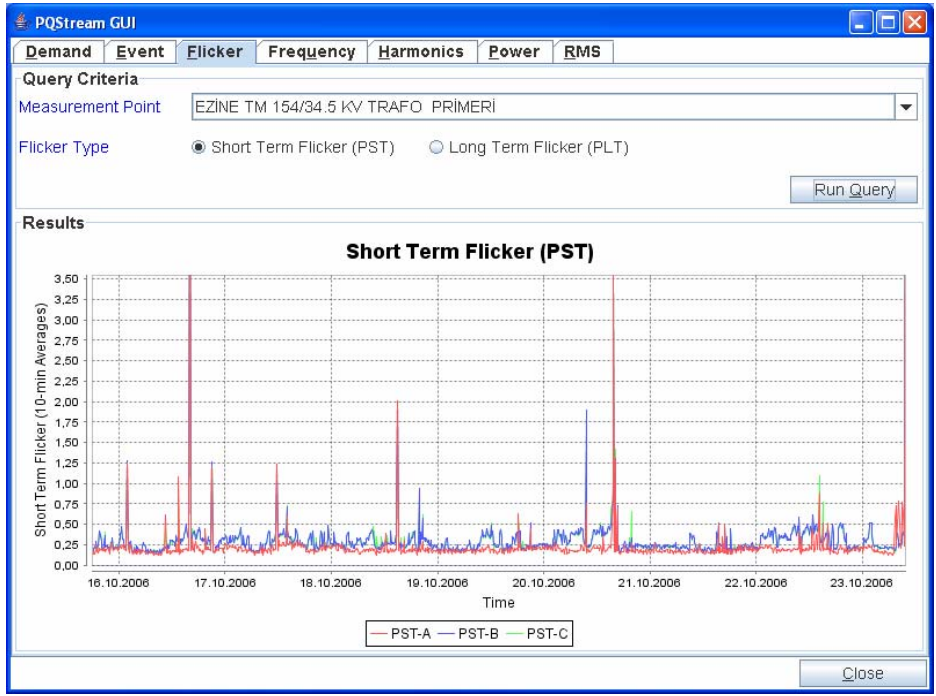

**Fig. 6.** PQStream GUI (with Flicker Query Panel)

- Users can provide aggregation queries through PQStream GUI so that summary information can be obtained about PQ parameters. For instance, the aggregation query provided in the Event query panel of the GUI in Fig. 7 can be represented in SQL as follows:

```
select     sum(es.sag_count), sum(es.swell_count),
           sum(es.unbalance_count),
           sum(es.event_count), mp.load_type
from       event_stat es, measurement_point mp
where      es.measurement_point_id = mp.id
group by   mp.load_type
```

- We have used Apache XML-RPC [16], Apache's Java implementation of XML-RPC protocol, for the communication between the stream database and PQStream GUI for its simplicity and compactness. Apache's Tomcat [17] web server is used to deploy server side PQStream code.

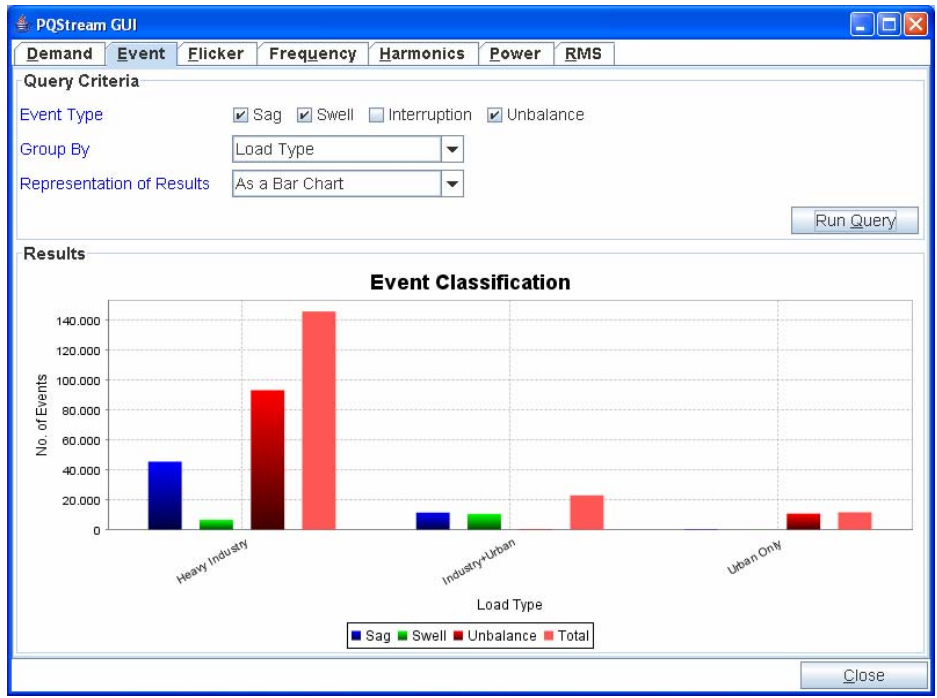

**Fig. 7.** PQStream GUI (with Event Query Panel)

## 4 Conclusion

Electrical PQ data is a time-evolving type of data and measuring it at high frequencies without any processing leads to unfeasibly large volumes requiring almost unbounded processing capabilities compared to other types of data stored in conventional relational database management systems. In this paper, we have described a data stream architecture for electrical PQ data, which, to our knowledge, is the first attempt to model PQ data as data streams, and shown its feasibility on real-world (field) PQ data.

    The main modules of PQStream architecture are a measurement module which processes continuous PQ data and computes averages according PQ standards, a stream database for storing the averages that the measurement module computed and finally a GUI for retrospective analysis and visualization of the stored PQ data. PQStream chooses not to store raw PQ data to lower the storage requirements of the acquired PQ data and uses some summary relations to speed up query processing. It supports aggregation queries over PQ data and with its proprietary GUI it enables users to access summaries of PQ data with relevant visualization facilities such as bar/pie charts and time-series graphs which are typical ways of presenting PQ data.

As further studies, we will employ data mining techniques on PQStream database such as classification and clustering algorithms to group those measurement points from which PQ data are acquired as well as sequence mining techniques to see the time-evolution of PQ problems. With the results of these data mining attempts on PQStream, experts of PQ domain will be able to take the necessary measures to detect and reduce PQ problems in electricity transmission systems.

**Acknowledgments.** This research and technology development work is carried out as a subproject of the National Power Quality Project of Turkey (Project No. 105G129, http://www.guckalitesi.gen.tr). Authors would like to thank the Public Research Support Group (KAMAG) of the Scientific and Technological Research Council of Turkey (TÜBİTAK) for full financial support of the project.